\documentstyle[12pt]{article}
\newcommand{\be}{\begin{equation}}
\newcommand{\ee}{\end{equation}}
\newcommand{\bea}{\begin{eqnarray}}
\newcommand{\eea}{\end{eqnarray}}

\begin{document}
\begin{titlepage}

\begin{flushright}
CERN-TH. 6711/92
\end{flushright}
\vspace{20 mm}

\begin{center}
\huge Black Hole
\vspace{1 mm}
\huge Uncertainties
\end{center}

\vspace{10 mm}

\begin{center}
Ulf H. Danielsson   \\
Theory Division, CERN, CH-1211, Geneva 23, Switzerland.
\end{center}

\vspace{20 mm}

\begin{center}
{\large Abstract}
\end{center}
\begin{flushleft}
\small In this work the quantum theory of two dimensional dilaton
black holes is studied using the Wheeler De Witt equation. The solutions
correspond to wave functions of the black hole. It is found that
for an observer inside the horizon, there
are uncertainty relations for the black hole mass and a parameter
in the metric determining the Hawking flux.
Only for a particular value of this parameter,
can both be known with arbitrary
accuracy. In the generic case there is instead a relation which is very
similar to the so called string uncertainty relation.
\end{flushleft}

\vspace{10mm}
\begin{flushleft}
CERN-TH. 6711/92 \\
November 1992
\end{flushleft}
\end{titlepage}
\newpage
\section{Introduction}

The latest developements in the study of two dimensional quantum black holes
have
 so far not given a final answer to the important question of whether or not
quantum coherence is lost in the presence of black holes. Perhaps this is not
too surprising. The treatments have so far been basically semi classical
and it is not clear whether all subtle issues about back reaction have been
properly taken into account. On the other hand, the fact that quantum
coherence is still unproven despite the huge amount of work done, might be
an indication that it may indeed be lost. At least in field theory
without strings.

A side effect of these developements is some revived interest in the
analysis of
precisely what is needed to save coherence, and the consequences of such
solutions \cite{harstro,presk}.
This area has been remarkably neglected in the past.

In the next section I will review what the problem exactly is, and some
ways out.
After this general overview, I will sketch the semi classical approach
to two dimensional quantum black holes. Then I will proceed with
an attempt
to improve these results by a study of
the Wheeler De Witt equation for the
system. This will take us on a journey through the horizon and into
the black hole.
I will illustrate some quantum mechanical uncertainty
relations obeyed by the black hole. In particular I will argue that
this has important consequences for the metric parameter determining
the Hawking flux.
Unless this has a specific expectation value, it can not be measured with
arbitrary accuracy by an observer inside the black hole.
Furthermore, in the generic case
the black hole mass can not
be measured, by the same observer, with an accuracy
better than on the order of a Planck mass.
This will be the main conclusion of this paper.

\section{The Problem of Quantum Coherence}

There is a well known example of loss of quantum coherence which do not
cause any trouble. This is the case of an open system which is part of
some larger, closed, system. When we want to make predictions about the
open system we should sum over all possible outcomes in the rest of
the system which occur together with
desired events in the open system for which we would like to calculate
probabilities. This sum, or trace, necessarily implies that it is impossible
to assign definite wave functions to the open system. It's only through the use
of density matrices which we can describe the time evolution. The states are
mixed rather than pure,
i.e one can only assign probabilities to which wave function we really
have.

One might add that this is very much similar to another well known example
where ordinary quantum mechanics is not applicable, i.e. the case of
measurements and the collapse of the wave function. After a measurement
we do not know {\it which} wave function we have. We only know that it
is one of the eigen functions corresponding to the measurement. We can
however assign probabilities to the different cases. So, after a
measurement the system is in a {\it mixed} state. Only after we have
checked the result of the measurement can we really know the true answer.

It is important to realize that this latter example taken at face value is
in fact a violation of quantum mechanics. Since, according to the unitary
time evolution of the Schr\"{o}dinger equation, pure states can not
evolve into mixed. Common ways out of this dilemma, is to claim that all
measurements involve open systems, and that this fact is
not properly taken into account.
The situation would then be more like the first case, where indeed quantum
mechanics holds along as we treat the whole closed system. That this really
works, remains unproven.

Enter black holes. Black holes are clearly a good way of supplying open
systems. Since the information hidden inside a black hole is inaccesable
to the world outside, this outsideworld may be thought of as an open system.
This is not too bad. The real trouble starts when we allow black holes to
Hawking evaporate. Then the black hole may simply disappear taking all its
information with it. Then there is nothing left on which we can blame the
loss of unitarity in our open system. Hence there is a {\it true},
rather than apparent violation of quantum mechanics.

Now, how do we avoid loss of quantum coherence?
The most conservative way out is to claim
that the Hawking radiation is in fact carrying the information.
Afterall, Hawking's original derivation, \cite{hawk}, is based on the
assumption that the radiation and the infalling matter do not interact.
Clearly this is not completely true. This is essentially the
observation made by t'Hooft, \cite{hooft}. Hence, the Hawking radiation
may be far from thermal and in fact capable of saving the purity.
There are two possiblities in this category. Either there is a continuos
leakage of information through Hawking radiation, or the information is
not returned until the very end of the evaporation process.
In the first case,
the Hawking radiation needs to
make a copy of the (quantum)
information contained in the infalling matter {\it before} it crosses the
horizon. But this means that there are very dangerous correlations
between the inside and the outside of the black hole. Precisely such
correlations which threatens quantum coherence. The trace which we are
forced to take over the inside, in its turn forces a partial collaps
of the outside world wave function through these very correlations.
Hence our attempt to save quantum coherence is the seed of its own
failure.

To get around this one could either assume that there never were nor ever will
be any correlations what so ever. But this merely means that we do not
know what we threw into the black hole in the first place. One could also
imagine, as in \cite{suss}, that {\it all}
information is reflected before it
reaches the horizon. No copy will ever enter and quantum coherence is
saved. But this implies the surprising conclusion that
it is impossible to
experience falling through a horizon. This clearly requires exceptional
quantum gravity effects to take place at the horizon. Since the horizon,
usually, is not thought to be a particularly dramatic place one is
inclined to view this argument as a ``reductio in absurdum" and rather
abandon quantum coherence.

According to the second possiblity, the information will cross the horizon,
but reemerge late in the evaporation process. How is this possible?
To understand this, the concept of a ``horizon" in an evaporating black hole
must be clarified.
There are two different horizons, the event horizon and the apparent horizon.
If one is inside an event horizon (which is a {\it global} construct), one
is doomed to hit the singularity. In the region inside an apparent horizon,
all directions in the lightcone point inwards, that is the radial coordinate
is timelike and the flow of time pushes you inward. In the case of the static
Schwarschild black hole the two horizons coincide. Now, if the black hole
is evaporating they do not agree anymore, in fact the apparent horizon is
shrinking and is timelike. This means that it is
possible to escape from within its realm.
However, since the apparent horizon is very nearly null, the escape
has to wait until very late in the evaporation process. The true
event horizon, on the other hand, might typically be of
Planck scale or even nonexistent.

There are several problems with this scenario. One is that it denies any
meaning to the black hole
entropy concept. Since, according to the Bekenstein formula,
the entropy is decreasing with the
mass, one would expect a continuos leakage of information,
like in the first case.
Here we have, instead, a Planck scale object still containing
information equivalent to that of a star. Another problem is that
this possibility in practice is equivalent to the strange ``stable
remnant" hypothesis. That is, with only a Planck mass worth of energy,
it will take a horrendous amount of time before a stellar amount of
information
can be transmitted. Since these remnants have so many different
states, there is a huge phase space available for there production,
which will cause problems even for low energy physics.

A remaining proposal is quantum hair. That is, it is assumed that there
exists a {\it huge} number of conserved quantities coupled to long
range gauge fields. Hence their conservation is guaranteed through
gauge invariance. Therefore not even black holes can break them. As
emphazised in \cite{presk}, this leads to a very strange world view.
Effectively, {\it all} different wave functions are in different
superselection sectors. This means that in such a world the superposition
principle loses its usual
meaning. Whether this state of affairs can remain
hidden from view in some low energy effective theory is not clear.
This is however claimed in \cite{mav}.

None of these possibilities are very appealing, clearly much work remains to
be done before the consequences are clarified. If nothing else, the
study of the two dimensional quantum black hole might help formulating
the correct questions. Afterall, all the conceptual problems of
quantum gravity (except the non renormalizability) are there.
Issues like the meaning of time and what the observables are in a
fluctuating geometry.
In this paper we will take a brief look at the latter,
finding some uncertainty relations.

After these general remarks, let us take a look at the two dimensional
quantum black hole in pursuit of the answers.

\section{The Outside View}

\subsection{Two Dimensional Quantum Black Holes}

In this section I will review some of the basic steps in solving the
two dimensional dilaton black hole, \cite{CGSH}.

The starting point for studies of the two dimensional dilaton black hole
is the string inspired action
\be
S = \frac{1}{4\pi} \int d^{2}x \sqrt{-g} \left[ e^{-2\phi}
(R + 4 (\nabla \phi )^{2} +4\lambda ^{2}) -\frac{1}{2}
\sum _{i=1}^{N} (\nabla f_{i})^{2} \right]
\ee
where the $f$'s are $N$ matter fields. From the conformal anomaly one
also gets a first quantum correction of Liouville type
\be
\kappa ( \partial _{-} \rho \partial _{+} \rho + \rho
\hat{R} )    .
\ee

{}From this, then, equations of motions and constraints can be derived which
allow for both stationary (Hartle-Hawking) and evaporating (Unruh) black
holes.

As observed in \cite{alwis,bilal,stro}, one also has the
freedom to add infinitely many
different extra terms which vanish in the weak coupling limit, i.e. far
from the black hole. Some of these choices are better than others, in the
sense that they lead to solvable models. In particular, there is a set of
theories which by some redefinitions of the fields can be brought onto a
familiarly looking form
$$
\frac{1}{2\pi} \int d^{2}x \left[ \mp \partial _{-} X \partial _{+} X
\pm \partial _{-} Y \partial _{+} Y \mp \frac{1}{2} \sqrt{\frac{
|\kappa |}{2}} Y \sqrt{-\hat{g}} \hat{R} +2\lambda ^{2}
e^{\sqrt{\frac{2}{|\kappa |}} (\mp X +Y)} \right]
$$
\be
 + matter.                 \label{verkan}
\ee
The exact choice of theory in this set is
given by the relation between $(X,Y)$ and $(\rho ,\phi )$. {\it One}
such, particularly simple, choice was suggested in \cite{suss}
$$
X= -\sqrt{\frac{2}{|\kappa |}} e^{-2\phi } \pm
\sqrt{\frac{|\kappa |}{2}} \phi
$$
\be
Y = \sqrt{2|\kappa |} \rho \mp
\sqrt{\frac{2}{|\kappa |}} e^{-2\phi }-
\sqrt{\frac{|\kappa |}{2}} \phi .            \label{transf}
\ee
Although the following treatment do not depend on this particular
choice, it is good to have it in mind as an example.

In the above expressions, where $\kappa = \frac{24-N}{6}$,
the upper sign is for $\kappa >0$ while the lower is for $\kappa <0$.
To simplify the notation, the calculations will in the following
be presented
for the case $N>24$ only, the case $N<24$ is analogous.

A static black hole space time is now described by
$$
X= -\sqrt{\frac{2}{|\kappa |}} (\frac{M}{\lambda} +
e^{\lambda (\sigma ^{+}
-\sigma ^{-})}) +Q\lambda (\sigma ^{+} - \sigma ^{-})
$$
\be
\label{kanon}
Y=-X +\sqrt{\frac{|\kappa |}{2}} \lambda (\sigma ^{+} -\sigma ^{-})
\ee
in asymptotically flat coordinates, i.e. as $r=\frac{1}{2}(\sigma ^{+}
-\sigma ^{-}) \rightarrow \infty$ we
have $\rho \rightarrow 0$.
$M$ is the black hole mass and $Q$ is a parameter which determines the Hawking
flux in the corresponding evaporating solution.
This solves the equations of motion derived from (\ref{verkan}).
As we will see, a careful consideration of the physics of the constraints is
needed to fix $Q$. For reference we also write down the solution in
coordinates
$$
x^{+} = \frac{1}{\lambda} e^{\lambda \sigma ^{+}}
$$
\be
x^{-} = -\frac{1}{\lambda} e^{-\lambda \sigma ^{-}}    .
\ee
It is given by
$$
X = \sqrt{\frac{2}{|\kappa |}} \lambda ^{2} x^{-} x^{+} +
Q\log |\lambda ^{2} x^{-} x^{+} | - \sqrt{\frac{2}{|\kappa |}}
\frac{M}{\lambda }
$$
\be
Y=-X          .
\ee

Using (\ref{transf}), one
easily recovers the classical solution
$$
e^{2\rho} =
\frac{1}{1+\frac{M}{\lambda} e^{\lambda (\sigma ^{-} -\sigma ^{+})}}
$$
\be
\phi = -\frac{1}{2} \log (\frac{M}{\lambda} + e^{\lambda (\sigma ^{+}-
\sigma ^{-})})
\ee
far away from the black hole.

For later use I will write down the solution in a more general gauge:
$$
X=B+C(\sigma ^{+} - \sigma ^{-}) - \lambda ^{2} \sqrt{\frac{|\kappa |}{2}}
\frac{1}{A^{2}}
 e^{\sqrt{\frac{2}{|\kappa |}}
\left[ A(\sigma ^{+} - \sigma ^{-})+ D \right] }
$$
\be
Y=-X+A(\sigma ^{+} -\sigma ^{-}) +D . \label{klos}
\ee
This is related to (\ref{kanon}), which I will call the ``canonical form",
by suitable coordinate transformations.
Later on, in section 4, it will be essential to consider this more
general expression.  With this in mind,
let us now find the necessary transformations.
Under coordinate transformations $X$ transforms as a scalar since it only
involves the dilaton $\phi$. $Y$, on the other hand, transforms like
\begin{equation}
Y \rightarrow Y+ \sqrt{\frac{|\kappa |}{2}} \log (\partial _{+} \sigma ^{+}
\partial _{-} \sigma ^{-})
\end{equation}
under $\sigma ^{\pm} \rightarrow \tilde{\sigma} ^{\pm}$, since it involves
a term $\sqrt{2|\kappa |}\rho$.

For convenience define
$$
A=a\sqrt{\frac{|\kappa |}{2}} \lambda \nonumber
$$
\be
C=c\sqrt{\frac{|\kappa |}{2}} \lambda    .
\ee
Then make the coordinate transformation
\begin{equation}
\sigma ^{\pm} =
\frac{1}{a} \tilde{\sigma}^{\pm} + \beta _{\pm}   \label{ktr}
\end{equation}
where
\begin{equation}
\beta _{+} -\beta _{-} = \frac{2\log a}{a\lambda } -\frac{D}{A} .
\end{equation}
We then get, (dropping the $\sim$'s), (\ref{kanon}) with
\be
M=-
[\sqrt{\frac{|\kappa |}{2}}B +\frac{c}{a} |\kappa | \log a -\frac{c}{a}
\sqrt{\frac{|\kappa |}{2}}D]\lambda
\ee
\be
Q= \frac{c}{a} \sqrt{\frac{|\kappa |}{2}}   \label{qekv}   .
\ee
The only undetermined parameters are now
$M$, the mass of the black hole, and $Q$.

\subsection{Evaporation Rates}

So far we have been studying static solutions, this will also be the case in
later sections. To find the proper meaning of the parameter $Q$, however, it
is useful to consider an evaporating solution. Such a solution is provided
in \cite{suss,alwis} by
$$
X = \sqrt{\frac{2}{|\kappa |}} \lambda ^{2} x^{-} x^{+} +
Q\log |\lambda ^{2} x^{-} x^{+} | + \sqrt{\frac{2}{|\kappa |}}
\frac{M}{\lambda x_{0}^{+}}(x^{+} - x_{0}^{+} )
\Theta (x^{+} - x_{0}^{+})
$$
\be
Y=-X          .
\ee
$x_{0}^{+}$ is the position of the infalling matter worldline.
In \cite{alwis} it is shown that $M$ is the Bondi mass when the black hole
is formed, and that $Q$ is proportional to the Hawking flux.

While having $M$ as a free parameter is precisely what we want, one would
rather see that $Q$ had some prefered value. As we have seen, however,
neither the equations of motion nor the constraints give us any guide
lines. It seems as if we have to make the choice on physical grounds.
Several alternatives have been proposed.

{\bf I}. $Q=-sign(\kappa ) \frac{1}{2} \sqrt{\frac{|\kappa |}{2}} $.

This is the unique value where the classical $M=0$ linear dilaton flat space
remains an exact
solution in the quantum corrected theory. The disadvantage is a
negative flux of Hawking radiation for $N<24$.

{\bf II}. $Q= \sqrt{\frac{2}{|\kappa |}} \frac{13}{12} $.

This is advocated in \cite{alwis}.
It can be motivated as follows. The energy-
momentum tensor consists of three parts:
\begin{equation}
T=T^{\rho ,\phi} + T^{M} + T^{gh}
\end{equation}
each transforming with a conformal anomaly. $26-N$,
$N$ and $26$ respectively.
The conformal anomaly of the gravity-dilaton sector
is by construction such
that the total anomaly vanishes. Let us now assume that the Hawking radiation
is carried {\it only} by $\rho , \phi$ and matter. Clearly, the flux must then
be proportional to $26-N+N=26$.

{\bf III}. $Q=\sqrt{\frac{2}{|\kappa |}}\frac{N}{24} $.

This is suggested in \cite{bilal} (note added).
The argument is now that only {\it matter}
should contribute. Hence the flux is $\sim N$. $\rho , \phi$ should not be
included since there are no propagating degrees of freedom in this sector.

However, this does not exclude contributions to the Hawking radiation. The
presence of Hawking radiation is simply a statement about the values of
the components of the energy-momentum tensor. These can be nonzero even without
any propagating degrees of freedom. ``No propagating
degrees of freedom"
simply means that there are no wavelike excitations which can be used to
transmit information or energy {\it relative} to what is already present.
Hence alternative
II can not be excluded on these grounds.

In section 4 I will reexamine these questions in a hopefully more
transparent setting and also
suggest some quantum properties which furthermore
distinguishes between these choices.

The parameter $Q$ will, by abuse of notation,
in the following frequently be refered to as ``the Hawking
flux". This in spite of the fact that we will be dealing with a
static black hole, and worse still, the space time {\it inside}
of the horizon. It is hoped that this will not lead to any confusion.

\section{Into the Black Hole}

\subsection{Canonical Quantization Inside a Black Hole}

In the ordinary $c=1$ quantum-gravity theory it is well known that the
mini superspace approximation is exact. The Wheeler De Witt equation for
the wave function of a loop inserted on the
world sheet coincides with the,
presumably correct, matrix model equation (see e.g. \cite{moore}).

Inspired by this, I will try to use the same approach for the
action (\ref{verkan}).
I will restrict myself to static solutions, leaving the time dependent
evaporating solutions for future work. The static solutions depend only on
$r=\frac{1}{2}(\sigma ^{+}-\sigma ^{-})$, this simplifying assumption will be
used in the following. The corresponding assumption in standard $c=1$ is that
the Liouville field, $\phi _{L}$, is constant around the loop, i.e.
$e^{-\phi _{L}} \sim l$, where $l$ is the loop lenght. It is also assumed that
the loop is inserted at a fixed point in target space, i.e. the matter field
is also constant around the loop.

The quantization procedure outlined will use $r$ as parameter. This means
that it is not
applicable to the region outside of the black hole, where $r$ is spacelike.
There we should rather use time $t$. {\it Inside} the black hole, however,
the roles are reversed. There $r$ is timelike, and $t$ is spacelike. Hence
$r$ is the correct choice. For an observer inside the horizon, the
``static" black hole is not static at all. Indeed, it is precisely
due the flow of time (i.e. $r$ in this region) that the observer
inevitable will end up in the singularity.
Therefore we will leave the treatment of the outside world for now,
and plunge into the forbidding interior of the black hole.

The wave functions we need to consider are then
of the form
\be
\psi = \psi \left[X(t), Y(t), r\right].
\ee
Since I will use the simplifying
mini superspace approximation, i.e. independence of $t$, this reduces to
ordinary quantum mechanics, i.e. $\psi = \psi (X,Y,r)$.

So, the mini superspace approximation amounts to the simplification
$$
\partial _{-} X \partial _{+} X = \frac{1}{4} (\dot{X}^{2} -X'^{2})
\rightarrow -\frac{1}{4} X'^{2} \nonumber
$$
\be
\partial _{-} Y \partial _{+} Y \rightarrow -\frac{1}{4} Y'^{2} .
\ee
{}From the action (\ref{verkan}), we then get the
canonical momenta
$$
P_{X} =-\frac{1}{2} X'
$$
\be
P_{Y} = \frac{1}{2} Y'
\ee
and the Hamiltonian is
\begin{equation}
{\cal H}
= P_{X}^{2} - P_{Y}^{2} +2 \lambda ^{2}e^{\sqrt{\frac{2}{|\kappa |}}
(X+Y)} .\label{ham}
\end{equation}
If we insert classical solution (\ref{klos}) we find
\begin{equation}
E=A(2C-A)     ,
\end{equation}
which implies
\begin{equation}
\frac{c}{a} = \frac{1}{2} + \frac{E}{2A^{2}}  .  \label{pjutt}
\end{equation}
For clarity I want to point out that in the case $N<24$, the signs are
reversed, hence
\begin{equation}
\frac{c}{a} =-\frac{1}{2} - \frac{E}{2A^{2}}  .
\end{equation}
This means that, using (\ref{qekv}),
case I in section 3.2 is special in the sense that it
corresponds either to $E=0$ or $A \rightarrow \infty$.

The quantum version of the canonical momenta are
$$
P_{X} =-i\lambda \frac{\partial}{\partial X}
$$
\be
P_{Y} =-i\lambda \frac{\partial}{\partial Y}   .
\ee
The $\lambda$ is inserted on dimensional grounds.
Consequently we obtain the
Schr\"{o}dinger equation
\begin{equation}
\lambda ^{2} [\frac{\partial ^{2}}{\partial Y ^{2}} -
\frac{\partial ^{2}}{\partial X^{2}} + 2 e^{\sqrt{\frac{2}{|\kappa |}}
(X+Y)}] \psi (X,Y) = E \psi (X,Y) .\label{schr}
\end{equation}
For the moment I will leave $E$ free. Some proposals how to fix it,
will be
given in section 4.2.
It is convenient to change variables to
$$
Z_{-} =\sqrt{\frac{2}{|\kappa |}} (Y-X) \nonumber
$$
\be
Z_{+} =\sqrt{\frac{2}{|\kappa |}} (Y+X)   .
\ee
In these variables the Schr\"{o}dinger equation becomes
\begin{equation}
\frac{\partial ^{2}}{\partial Z_{+} \partial Z_{-}} \psi
+\frac{|\kappa |}{4} e^{Z_{+}} \psi
= \frac{|\kappa |}{8\lambda ^{2}} E \psi .
\end{equation}
One solution is easy to find. This is
\begin{equation}
\psi \sim e^{i\alpha Z_{+} +i\beta Z_{-} +
\frac{i|\kappa |}{4\beta} e^{Z_{+}}}    \label{vag}
\end{equation}
with $\frac{|\kappa |}{8\lambda ^{2}} E = -\alpha \beta$.
One might note that this solution is such that WKB is exact. It is
essentially just a plane wave.

Before we can proceed, we must make sure that we are using coordinates
appropriate for the interior of the black hole. The
coordinates $x^{\pm}$ may be used everywhere, but let us rather find
the analog of the $\sigma ^{\pm}$ coordinates. These are
clearly just good for the outside of the $x^{-}x^{+}=0$ lightcone.
Asymptotic infinity is reached as
$r \rightarrow +\infty$, while the $x^{-}x^{+}=0$ lightcone is
where $r \rightarrow -\infty$.
The interior version needs a change of sign, we put
$$
x^{+} = \frac{1}{\lambda} e^{\lambda \hat{\sigma}^{+}}
$$
\be
x^{-} = \frac{1}{\lambda} e^{-\lambda \hat{\sigma}^{-}}    .   \label{inne}
\ee
In the $\hat{\sigma}^{\pm}$ coordinates the $x^{-}x^{+}=0$ lightcone
is again where $r \rightarrow - \infty$,
but as $r$ increases we now move inwards reaching the singularity at some
finite value of $r$. A well posed initial value problem is then obtained
by prescribing values for the fields on some space like slice of
constant $r$. As $r \rightarrow - \infty$ this asymptotes to
the $x^{-}x^{+}=0$ lightcone.

The analog of (\ref{klos}) (and (\ref{kanon})), is, through
a coordinate transformation, seen to be obtained by a simple substitution
$D \rightarrow D + \sqrt{\frac{|\kappa |}{2}} \log (-1)$. In $X$ this
amounts to a change of sign of the exponential term, while in $Y$ the shift
can be absorbed in a redefinition $e^{2\rho} \rightarrow -e^{2\rho}$.

It should be noted that there is a region {\it in between}
the horizon, as defined by $\partial _{+} \phi =0$, i.e.
\begin{equation}
x_{h}^{-} = -\sqrt{\frac{|\kappa |}{2}} \frac{Q}{\lambda ^{2}}
\frac{1}{x_{h}^{+}}
\end{equation}
for the static black hole, and the $x^{-}x^{+}=0$ lightcone were
the original $\sigma ^{\pm}$ coordinates must be used. In a Penrose
diagram this region must be considered with care,
tipping the lightcones the right way.
If one is careful with one's definitions, most formulae I will be
using are very much the same both inside and outside of the horizon,
and inside and outside of the $x^{-}x^{+}=0$ lightcone. But,
of course, it is only {\it inside} of the horizon
where the quantum
interpretation clearly makes sense.

To make contact with the classical case, we compute the momenta
$$
X' = -2P_{X}=2i\lambda \frac{\partial}{\partial X} \log \psi \nonumber
$$
\be
=-
2\lambda [ (\alpha -\beta )\sqrt{\frac{2}{|\kappa |}} +\frac{1}{2\beta }
\sqrt{\frac{|\kappa |}{2}}e ^{\sqrt{\frac{2}{|\kappa |}} (X+Y)} ]
\ee
and
\begin{equation}
Y'=2P_{Y}=-X'+4\beta \sqrt{\frac{2}{|\kappa |}} \lambda   .
\end{equation}
$\alpha$ and $\beta$
must be real to give a real classical solution.
By comparing with the classical solution we find, in the notation of
(\ref{klos})
$$
A=2\beta \lambda \sqrt{\frac{2}{|\kappa |}} \nonumber
$$
\be
C=(\beta -\alpha )\lambda \sqrt{\frac{2}{|\kappa |}}       ,
\ee
while neither $B$ nor $D$ is determined to have any specific value. Here
we see the uncertainty principle at work, the consequences will be
analysed in section 4.3.

It should be noted that inside of the horizon it is very
tricky to define the
vacuum since there is no asymptotically flat region. For the present
discussion, however, this question is not important.

\subsection{Wave Functions and Vertex Operators}

As in the $c=1$ case the wave functions we have obtained are related to
vertex operators. In string theory they create particles living in the
target space. Here we do not have that interpretation since the two
dimensional ``world sheet" {\it is} our space time. Rather, the vertex
operators might
create new universa in the sense of third quantization.

Let us see if we can mimic the vertex operator concept in the black hole
case. If we succeed, this will allow us to fix the value of $E$ in the last
section. Unfortunately, as I will point out later on, this will not help
us in fixing $Q$ and I will try to show why. The results of the next, and
most important section, are hence largely independent of whether or not we
can fix $E$.

The natural context for a vertex operator is Euclidean radial quantization.
The surfaces of quantization are the circles $z\bar{z} = const$. ``Initial
conditions" are provided by some vertex operator sitting at the point
$z\bar{z}=0$. The Minkowsky counterpart (for noncompact space) is
quantization using the spacelike hyperbolas $x^{+}x^{-} = const$.
Initial conditions must now be specified on the full future
lightcone $x^{-}x^{+}=0$ ($x^{-},x^{+} \geq 0$).
The noncompact space dimension is a major technical difficulty, this
means that the ordinary string theory tools, i.e. mode expansions
etc. are not directly applicable. Below I will be restricting the
attention to the mini superspace approximation, hopefully, therefore,
these questions should not affect the conclusions.

With the Minkowsky/Euclidean differences (and the non compactness
of space) in mind, let us try to exploit the analogy.

A natural condition to impose on the vertex operator $\cal T$ is then the
string theory
physical state condition
\be
T_{++} |{\cal T}> =\frac{1}{(x^{+})^{2}} |{\cal T}>  \label{kond1}
\ee
As we will see, there is no loss of generality in assuming units such that
the righthand side prefactor is $1/(x^{+})^{2}$ rather than, say,
some general $\eta /(x^{+})^{2}$, as long as $\eta \neq 0$.
Let us try this and investigate the consequences.
We also need some physical input.
Let us go to the asymptotically flat
$\sigma ^{\pm}$ coordinates, for the moment denoting the
energy momentum tensor in these coordinates by $\tilde{T}_{++}$. The physical
condition is that there are neither matter nor ghosts present. Hence
$\tilde{T}_{++}^{M}= \tilde{T}_{++}^{gh}=0$. In case of a static black
hole, one also has $\tilde{T}_{--}=\tilde{T}_{++}$. This needs some
clarification. Even in a static black hole, one would say that
there are contributions to e.g.
the energy momentum tensors
$\tilde{T}^{M}_{++}$ and $\tilde{T}^{M}_{--}$. Although they remain
equal so that there is no net energy flow.
These contributions correspond to vacuum polarizations which fall off
exponentially as one receeds from the black hole. As $r \rightarrow
\infty$
they both go to zero. Now, why should not these be included in the
conditions above? Clearly the vacuum is not empty as the conditions above
seem to indicate. The point is that these contributions are {\it
by definition} already included in $\tilde{T}^{X,Y}$. The correct
prescription, then, is to find coordinates associated with the desired
vacuum, in this case the asymptotically flat $\sigma ^{\pm}$
coordinates. Then assign values to $\tilde{T}^{M}$ (and $\tilde{T}^{gh}$)
corresponding to {\it classical} matter. In case of a forming and evaporating
black hole we would need to add, typically, a contribution $\sim \delta
(x^{+}-x_{0}^{+})$ to $\tilde{T}_{++}^{M}$. For the study of
the space time structure,  questions about what is the vacuum energy etc.
are largely irrelevant. The only thing which matters is the value of
$T^{X,Y}$.

Through the anomalous
transformations, we get $T_{++}^{M}= \frac{-N}{24(x^{+})^{2}}$ and
$T_{++}^{gh} =\frac{26}{24(x^{+})^{2}}$. Now, let us see what this
implies for
the classical value of $T_{++}^{X,Y}$, from which we can deduce e.g. the
values of $E$ and $Q$. This classical part is transforming with a
conformal anomaly $24-N$. In addition there is a quantum anomaly with
value $2$, giving a total $26-N$. When we want to read of the {\it classical}
part of $T_{++}^{X,Y}$ this should not be included. Adding all these
contributions, using the condition (\ref{kond1}), we find
\be
T_{++}^{X,Y} |{\cal T}>= \frac{N}{24(x^{+})^{2}} |{\cal T}>
\ee
This can be written as
\begin{equation}
\lambda ^{2}
(-\frac{1}{2} \frac{\partial ^{2}}{\partial X^{2}} + \frac{iQ_{B}}{2}
\frac{\partial}{\partial Y} + \frac{1}{2} \frac{\partial ^{2}}{\partial Y^{2}}
+e^{\sqrt{\frac{2}{|\kappa |}}(X+Y)}
){\cal T}=\frac{N}{24(x^{+})^{2}} {\cal T}
\end{equation}
where the background charge $Q_{B}$ is
\begin{equation}
Q_{B}
=2\sqrt{\frac{\kappa }{2}}
\end{equation}
For
$N=0$ this is $2\sqrt{2}$, the well known value for the $c=1$ theory,
(in units where $\alpha '$ =2).

The vertex operator $\cal T$ is related to the wavefunction $\psi$ through
\begin{equation}
{\cal T}(x^{-},x^{+}) \rightarrow
e^{-\frac{iQ_{B}}{2} Y} \psi (\hat{\sigma}^{-},\hat{\sigma}^{+})
\end{equation}
The extra exponential in front of $\psi$ is due to the singular
change of variables (\ref{inne}), there is a $\delta$ function
contribution to the curvature in the action, see e.g. \cite{shenk}.
{}From this it follows that
$\psi$ solves (\ref{schr})
(note that ${\cal H}= T_{++} +T_{--} -2T_{-+}=2T_{++}$)
with
\begin{equation}
E=2\lambda ^{2}(\frac{N}{24} -\frac{Q_{B}^{2}}{8}) =
2\lambda ^{2}.
\end{equation}
For $N<24$ the sign is reversed.
This is then the value of $E$, dictated by the physical state condition
which we should use in the Schr\"{o}dinger equation (\ref{schr}).

It is simple to check that
$T^{X,Y}_{++}$,
due to its anomalous transformation
properties,
transforms like $\frac{N}{24(x^{+})^{2}} \rightarrow
\frac{N}{24} -\frac{N-24}{24} =1$ as it should.

Now, from this one might be tempted to think that we
have fixed $Q$ to $\sqrt{\frac{|\kappa |}{2}} Q=
\frac{N}{24}$, i.e. alternative III.
This is however not true. To know
what the value for $Q$ is, we must also know the semiclassical values for
the metric and dilaton fields. These are contained in the wave function
$\psi$. Given a wavefunction, the coordinates $\hat{\sigma}^{\pm}$ will
in general
correspond to a coordinate system in which the $X$ and $Y$ fields are
of the form (\ref{klos}), (interior version).
Therefore a coordinate transformation
(\ref{ktr}) might be needed.
Recall the expressions (\ref{klos}).
In agreement with (\ref{pjutt}) we get
\be
\sqrt{\frac{|\kappa |}{2}} Q=
\frac{N-24}{24} +\frac{1}{a^{2}}
\ee
(for boths signs of $\kappa$). A general prefactor $\eta /(x^{+})^{2}$
(with $\eta \neq 0$) changes this to
$\frac{N-24}{24}+\frac{\eta}{a^{2}}$, hence no qualitative difference.
Note that the Schwarzian derivative of
(\ref{ktr}) is zero, as required for consistency. The string theory
analog is that there are many different tachyons, with various energies
and momenta which solve the physical state condition.

So, our fixing of $E$ has not really accomplished much, {\it but} it has
focused our attention on the fact that the value of the Hawking flux, $Q$
(and also the black hole mass) are specified through the wave function
$\psi$. Clearly one can imagine having superpositions of such
$\psi$'s, all corresponding to the same fixed eigen value $E$, but
each requiring a different coordinate transformation to extract the
right values for $Q$,
yielding different degrees of uncertainty for these parameters.
In the next section we will explore these issues in greater depth.

The condition (\ref{kond1}) is not the only one which deserves attention.
In string theory there is also the unique
$SL(2,C)$ invariant vacuum which do
not solve the physical state condition. Instead the $L_{0}$ constraint is
shifted to $L_{0}=0$. Hence one might try
\be
T_{++} |{\cal T}> =0    \label{kond2}
\ee
By repeating the above argument, one finds $\tilde{T}_{++}^{X,Y}
=\frac{N-24}{24}$, $E=0$ and $\sqrt{\frac{|\kappa |}{2}} Q= \frac{N-24}{24}$.
This is alternative I. In this case $Q$ is uniquely determined, there is no
dependence on $a$.

In this section we have seen how to translate the issue of choice of
vacuum in the quantum black hole to string theory language.
Let us now in more detail investigate the physical consequences.

\subsection{Black Hole Uncertainties}

I now come to the central part of the discussion.
In section 4.1 we saw an example of a black hole wave function,
i.e. (\ref{vag}). Some
of the parameters in (\ref{klos})
were exactly determined by the wavefunction in the example,
i.e. $A$ and $C$, while others were not, i.e. $B$ and $D$. The reason
is the uncertainty principle.
By considering the conjugation rules for the fields and their momenta, the
following uncertainty relations can be shown to hold at best
\be
\Delta C \Delta B  \sim  \frac{\lambda }{2} \hspace{3mm}
\Delta C \Delta D \sim \frac{\lambda }{2}   \hspace{3mm}
\Delta A \Delta B \sim \frac{\lambda }{2}    \label{konj}
\ee
The pairs $(A,D)$, $(D,B)$ and $(A,C)$ on the other hand, obey no
such restrictions.
Clearly this will have consequences for how well determined the
black hole mass is for a given wave function.
I will focus on case (\ref{kond1}).

The wave functions must be chosen, as we have seen, as eigen functions of
the Hamiltonian (\ref{ham}) with a certain
eigen value $E$. This
means that the combination $A(2C-A)$ must be completely fixed, and hence
the errors in $A$ and $C$ {\it correlated} through
\be
\Delta C = \frac{A-C}{A} \Delta A  .
\ee
Respecting
this condition we can compute the error in $M$ to be
$$
\Delta M \sim
\lambda \sqrt{\frac{|\kappa |}{2}} \Delta B
+\frac{c}{a} \sqrt{\frac{|\kappa |}{2}} \lambda \Delta D
$$
\be
+|\sqrt{2|\kappa |} \left[ (1-\frac{2c}{a})\log a +\frac{c}{a} \right] -
(1-\frac{2c}{a})D| \frac{\Delta A}{a}  .
\ee
We have
\begin{equation}
\Delta D \sim \frac{\lambda}{2\Delta C} \sim \frac{A}{2(A-C)}
\frac{\lambda}{\Delta A}    .
\end{equation}
Also
\begin{equation}
\Delta B \sim  \frac{\lambda}{\Delta A}  .
\end{equation}
(assuming $A>C$).
Hence
\begin{equation}
\Delta M \sim f(a,c)\frac{1}{\Delta A} + g(a,c,D) \Delta A  \label{osak}
\ee
where
$$
f(a,c) =
\lambda ^{2} \sqrt{\frac{|\kappa |}{2}}
(1+|\frac{c}{2(a-c)} |)
$$
\be
g(a,c,D) =
|\sqrt{2|\kappa |} \left[ (1-\frac{2c}{a}) \log a +\frac{c}{a} \right] -
(1-\frac{2c}{a})D | \frac{1}{a} .
\end{equation}
If we now try to minimize the error in the black hole mass $M$ with respect
to $\Delta A$, we find  $\Delta A \sim \sqrt{\frac{f}{g}}$ and
a {\it nonzero} minimal error in $M$, given by $\Delta M \sim 2
\sqrt{fg}$.

We should also note that the Hawking flux $Q$, which involves $c/a$ given
by
\begin{equation}
\frac{c}{a} = \frac{1}{2} + \frac{\lambda ^{2}}{A^{2}}
\end{equation}
also
has a finite minimal error, since
\begin{equation}
\Delta (\frac{c}{a})
= \frac{2\lambda ^{2}}{A^{3}} \Delta A .
\end{equation}
The only way to make the error in the Hawking flux disappear, and
still have a finite error in the black hole mass,
is to take $A$ to infinity. Then both the
error in the black hole mass and the error in the
Hawking flux will go to zero.
By suitable adjustments of the other parameters, one can still find wave
functions describing arbitrary black hole masses (now well determined).
{\it But} there is a price to be paid. When $A \rightarrow \infty$,
$\frac{c}{a}$ is driven to the value $1/2$.
This was alternative I in section 3.
Clearly, the $A \rightarrow \infty$ limit of the (\ref{kond1}) case,
is equivalent to the case (\ref{kond2}).

If we insist on $c/a \neq 1/2$, $\Delta M$ can still be zero for
some value of $D$, giving $g=0$. This means however that
$\Delta A \rightarrow \infty$ and the flux is totally unknown.
This corresponds to an uncertainty principle relating the mass $M$
and the flux $Q$.
The generic case, however, is that there is an absolute upper
limit on the accuracy in the black hole mass.
I will comment on this shortly.

Hence we conclude that if we want a world including a black hole with a
Hawking flux different from alternative I, we are free to have that, but
in that world the Hawking
flux can not be known with arbitrary accuracy, and typically not the
black hole mass either, (at least not by an observer {\it inside} the
black hole).

Let me for clarity repeat the argument again using less algebra. Let us
assume that a given wave function is peaked around a certain solution, for
simplicity already on canonical form. (It should be emphazised that there
is nothing special about the ``canonical form". It's just a convenient
step on the way when one wants to read off, e.g.,
the mass of the
black hole.) Furthermore, let us assume that the black hole mass is
arbitrary precise, i.e. $\Delta B =0$.
Through the uncertainty relations,
however, this implies that $A$ is completely unknown! That is, we can't
really be sure if we are in the canonical gauge anymore. Hence
the formula for the mass of the black hole is corrected
by an unknown amount.
It is not simply $B$ anymore. Hence our original hope is falsified, the
black hole mass is not well determined.

One may note that the relation (\ref{osak}),
is very similar to the string
uncertainty relation (see e.g. \cite{gross})
\begin{equation}
\Delta x \sim 1/E + E \label{strosak}
\end{equation}
which shows that there exists an effective minimal lenght, ($\sim$
Planck scale) in the (string) world.
Here the first term is quantum mechanical while the second one is due
to the
extent of the probe, i.e. the string.
{}From this point of view it is better to think of (\ref{osak}) in terms of
the Schwarzschild radius (if we imagine a similar relation to hold for
four dimensional black hole) rather than the black hole mass. The relation
(\ref{strosak}) then suggests an uncertainty in the Schwarzschild radius
of a black hole which agrees in form with (\ref{osak}). In the black
hole case the probe is the combined gravity-dilaton field.

We have been studying field theory, not strings. But in two
dimensions, quantum gravity is possible even without strings. Furthermore,
since the only propagating degree of freedom is the tachyon, string theory
is not too
different from field theory in two dimensions.
It is therefore quite remarkable that we find a relation so similar to
(\ref{strosak}).
\section{Conclusions}

I have tried to argue that the uncertainty principle plays an important role in
the study of the two dimensional dilaton black hole. Both the black hole
mass and the parameter determining the
flux of Hawking radiation are affected in the eyes of
a traveller venturing into the black hole. Only if the flux parameter
takes a
very specific value, $\frac{N-24}{24}$,
can both
be measured with arbitrary accuracy by the black hole explorer.
In string theory this corresponds to the $SL(2,C)$ invariant vacuum.
In the generic case, one obtains uncertainty relations
very similar to the
so called string uncertainty relation. This situation is obtained by
using the string physical state condition.

As far as the evaporation process is concerned the uncertainty relation
is not at all unwelcomed. If we assume a similar relation to hold even
in this case,
it means that a Planck mass black hole has the option
to disappear through quantum fluctuations. Hence the singularity
may cease to exist {\it before} it reaches the horizon (as seen
from the inside) and threatens to emerge like a naked singularity,
(as seen from the outside).

Another important problem which should be apparent from the analysis
is the choice of
coordinate for quantization, i.e. time.
This is especially
important close to a black hole horizon. If the geometry is fluctuating
how do we choose a timelike direction and spacelike surfaces?
Clearly this is a dynamical question.
Such questions must be answered before a true understanding of the
quantum black hole beyond the semiclassical picture
can be achieved, \cite{bo}.

The mini super space approach seems to be a powerful and intuitively
appealing tool for studying the quantum black hole. Clearly it would be
worthwhile generalizing this treatment to the case of an evaporating
black hole.

In the case of $N=0$, one has an even more powerful tool at disposal,
the matrix model. As noted in \cite{sakai},
the only difference is that the
cosmological term is replaced by the lowest momentum special tachyon,
the one with $SU(2)$ quantum numbers $(J,m)=(1/2,1/2)$. For an
introduction, see \cite{avhand}. Therefore, in principle, the $N=0$ case
should be exactly solvable. Not only with trivial space time topology,
but also including higher genus corrections. In our space time
language this corresponds to worm holes.

These issues will be left for future work.

\section*{Acknowledgements}

I would like to thank L. Alvarez-Gaume, T. Inami, K. Lee and
X. Shen for discussions. I would also like to thank N. Mavromatos
for explaining his work. Finally, I would like to thank
Bo Sundborg, Stockholm, for illuminating discussions on the
concept of space time in the presence of quantum mechanics.

\end{document}